# Local order in binary Ge-Te glasses – an experimental study

P. Jóvári[1], A. Piarristeguy[2], A. Pradel[2], I. Pethes[1], I. Kaban[3], S. Michalik[4], J. Darpentigny[5], R. Chernikov[6]

[1]Wigner Research Centre for Physics, Institute for Solid State Physics, H-1525 Budapest, POB 49, Hungary

[2]Institut Charles Gerhardt, UMR 5253 CNRS, CC1503, Université de Montpellier, Pl. E. Bataillon, F-34095 Montpellier cedex 5, France

[3]IFW Dresden, Institute for Complex Materials, 01171 Helmholtzstr. 20, 01069 Dresden, Germany

[4]Diamond Light Source Harwell Science and Innovation Campus, Didcot, Oxfordshire, OX11 0DE, UK

[5]Laboratoire Léon Brillouin, CEA-Saclay 91191 Gif sur Yvette Cedex France

[6]Deutsches Elektronen Synchrotron DESY, Notkestrasse 85, D-22603 Hamburg, Germany

Abstract

The structure of $Ge_xTe_{100-x}$ ($x$ = 14.5, 18.7, 23.6) glasses prepared by twin roller quenching technique was investigated by neutron diffraction, X-ray diffraction and Ge-K-edge X-ray absorption spectroscopy measurements. Large scale structural models were obtained for each composition by fitting the experimental datasets in the framework of the reverse Monte Carlo technique. It was found that the majority of Ge and Te atoms satisfy the 8-N rule. Simulation results indicate that Ge-Ge bonding is not significant for $x$ = 14.5 and 18.7. The shape and position of the first peak of the Ge-Ge partial pair correlation function evidence the presence of corner sharing tetrahedra already in compositions ($x$ = 14.5 and 18.7) where 'sharing' of a Te atom by two Ge atoms could be avoided due to the low concentration of Ge.

1. Introduction

Short range order of amorphous Ge-Te alloys has been investigated for decades by diffraction and extended X-ray absorption fine structure (EXAFS) measurements [1-8]. Some early studies suggested the so called 3-3 model (both Ge and Te are threefold coordinated) but later the 4-2 model became prevailing. However, some recent works raised doubts about the validity of the 8-N rule [9-11]. The precise determination of the coordination numbers of Ge and Te in amorphous Ge-Te alloys is a rather difficult task. The main problem is the overlap of the first peaks of $g_{GeTe}(r)$ and $g_{TeTe}(r)$ partial pair correlation function, which makes their separation - and the determination of corresponding coordination numbers - difficult. This was clearly illustrated by the neutron diffraction study of Ichikawa *et al.* [4] who found that the first peak of neutron diffraction total pair correlation function of $Ge_{20}Te_{80}$ glass could be fitted by a single Gaussian. From this they concluded that there are only Ge-Te bonds in the sample and the average number of Te atoms around Ge is $6.3 \pm 0.4$.

The situation can be improved by combining neutron diffraction, X-ray diffraction and Ge K-edge EXAFS data. According to our experience, the latter gives a very precise estimate of the Ge-Te peak position while diffraction datasets reduce the uncertainty of coordination numbers.

Detailed knowledge of the structure of binary Ge-Te glasses is useful in structural investigations of more complex telluride glasses used in infrared optics or information technology. A thorough experimental study may also inspire theoreticians by providing them with reliable structural models. The aim of the present study is the determination of short range order in Ge-Te glasses prepared by twin roller quenching using the above mentioned experimental techniques. The reverse Monte Carlo simulation technique (RMC) [12-15] will be used as a framework for combining the information content of different experimental datasets. This method is widely used to study the structure of liquids and glasses (see e.g. [16, 17]). The uncertainty of short range order parameters is also investigated. Though the glass forming region of the Ge-Te system is not too wide, comparison of the structure of different compositions helps in assessing the reliability and significance of our results.

2. Experimental

Bulk $Ge_xTe_{100-x}$ samples ($x$ = 14.5, 18.7, 23.6) were synthesized from a mixture of high-purity starting elements (Ge (Aldrich, 99.999%) and Te (5N+, 99.999%)). Four grams of the stoichiometric powders were placed in a silica tube and sealed under secondary vacuum ($10^{-5}$ mbar). The powder was melted at 660ºC with a 9ºC/h heating rate and kept at this temperature for 24 h for homogenization before being quenched in a salt-ice-water mixture. The obtained materials were then crushed into small pieces. The pieces were put in a quartz tube, having a hole at the bottom. They were further heated using a RF induction furnace, which allowed the temperature of the mixture to rise very rapidly. Pressure, induced via an Ar gas jet, was applied to the melt forcing small droplets through the hole and in between two rotating twin rollers. Quenched flakes (45-85 μm in thickness, a few cm$^2$ in area), released by the rollers, dropped into an Al collector. The twin roller quenching equipment was put in an Ar-filled glove box (for more details see Pradel *et al.* [18]).

The chemical composition was estimated by Electron Probe Micro-Analyser (EPMA) using a CAMECA SX-100 instrument with an acceleration voltage of 20 kV and a probe current of 1 nA. The amorphous nature was checked by X-ray diffraction using a PANalytical XPERT diffractometer. A Cu (K$\alpha$) source ($\lambda$ = 1.5406 Å) was used for the excitation with operating voltage of 40 kV and a beam current of 30-40 mA.

Neutron diffraction (ND) data were collected at the 7C2 liquid and amorphous diffractometer [19] of LLB (Saclay, France). Powdered samples were filled into vanadium sample holders with 5 mm diameter and 0.1 mm wall thickness. Wavelength, detector position and efficiency was determined by auxiliary measurements of Ni and V powders, respectively. The wavelength of incident neutrons was 0.724 Å. Raw data were corrected for background scattering, multiple scattering and absorption using standard procedures.

The X-ray diffraction experiment was carried out at the Joint Engineering, Environmental and Processing (I12-JEEP) beamline [20] at Diamond Light Source Ltd (UK). The size of the monochromatic beam was $0.3 \times 0.3$ mm$^2$. The energy of the incident beam, the sample-to-detector distance, the position of the beam centre and the tilt of the detector were determined by measuring a CeO$_2$ standard sample (NIST Standard Reference Material 674b) at different distances [21] The wavelength of the incident beam and the sample-to detector distance were 0.1255 Å (98.768 keV) and 336 mm, respectively.

Powder samples were filled into quartz capillaries with diameter of 1 mm and wall thickness of 0.01 mm. X-ray data were measured in transmission geometry by a large area 2D detector (Thales Pixium RF4343). The illumination time for a single diffraction pattern was 20 s, and 45 images were collected for each sample to improve statistics at high $k$-values. An empty quartz capillary was also measured at the same conditions and the resulting 2D image was subtracted from the sample images. Background corrected images were integrated into the $k$-space using the DAWN software [22]. Structure factors were obtained by correcting the integrated data for self-absorption, Compton scattering, fluorescence and multiple scattering using the PDFGetX2 program [23].

Corrected neutron- and X-ray diffraction structure factors are shown in Figure 1.

Ge K-edge EXAFS data were collected at beamline P65 of PETRA III synchrotron storage ring (Hamburg, Germany). Experiments were carried out in transmission mode. The monochromatic beam was obtained by a Si(111) double crystal monochromator. Powdered samples were mixed with cellulose and pressed into tablets. The amount of sample was chosen to get an absorption μt≈1.5 above the absorption edge. Intensities before and after the samples were measured with ionization chambers filled with Ar/Kr mixture and Kr, respectively.

The raw absorption spectra were converted to χ(k) curves using the VIPER program [24]. Raw, $k^3$-weighted χ(k) data signals were forward Fourier-transformed first into *r*-space using a Kaiser-Bessel window (α=1.5). The *k*-range of transformation was 2 Å$^{-1}$ -15 Å$^{-1}$. The *r*-space data were back-transformed using a rectangular window over 1.1 Å-2.8 Å (see Figure 2).

## 3. Simulation

Calculations were carried out by the rmcpp programme [13]. Simulation boxes contained 32000 atoms. Densities were determined by interpolating the molar volumes of amorphous Ge$_{15}$Te$_{85}$ [25] and Ge$_{50}$Te$_{50}$ [26]. Minimum interatomic distances (cut offs) were 2.35 Å and 2.55 Å for Ge-Te and Te-Te pairs, respectively. For Ge-Ge pairs 3.5 Å was used when Ge-Ge bonding was forbidden and 2.35 Å was applied when it was allowed. In these simulations only unphysically low coordination numbers were forbidden (0 for Te, 0 and 1 for Ge) but the total coordination number of atoms was not constrained otherwise.

Fits of the 'final' model of Ge$_{18.7}$Te$_{81.3}$ are shown in Figure 3.

## 4. Results and discussion

### 4.1. Ge-Ge bonds in Ge-Te glasses

To see whether Ge-Te glasses are chemically ordered, the occurrence of Ge-Ge bonds was tested first. It was found that for Ge$_{14.5}$Te$_{85.5}$ and Ge$_{18.7}$Te$_{81.3}$ the Ge-Ge cut off could be raised to 3.5 Å without changing the quality of fits. Therefore, in these compositions $N_{GeGe}$, the Ge-Ge coordination number, is below the estimated sensitivity of our approach (~0.3). In case of Ge$_{23.6}$Te$_{76.4}$ fits were improved by allowing Ge-Ge bonds in model configurations. Here the estimated value of $N_{GeGe}$ is 0.44 ± 0.3 (see Table 1). For a completely random 4-2 glass the Ge-Ge coordination number is given by the following equation [27]:

$$N_{\text{GeGe}}(x) = 4\frac{2x}{100+x} \qquad (1)$$

For $x$=23.6 $N_{GeGe}$ would be 1.53 according to the above equation. Our result therefore clearly indicates that twin roller quenched Ge-Te glasses cannot be described as random covalent networks.

### 4.2. Short range order parameters

Partial pair correlation functions $g_{ij}(r)$ (with $i, j$=Ge and Te) obtained by unconstrained simulation runs can be found in Figure 4, while coordination numbers and bond lengths are shown in Tables 1 and 2. Deviations of the total coordination numbers of Ge and Te from 4 and 2 are within the ± 4% range for all compositions. Again, it is to be emphasized that these values were obtained by *unconstrained simulations*, in which the total coordination number of Ge and Te were not forced to be 2 and 4. The uncertainty (range of deviation compatible with experimental datasets) of the coordination numbers was estimated by dedicated simulation runs, in which values given in Table 1 were forced to increase or decrease in steps of 5-10% and the quality of fits was monitored. The errors of $N_{TeGe}$ and $N_{TeTe}$ are usually not higher than 0.15 while the uncertainty of $N_{Te}$ is around 0.1 showing that changes of $N_{TeTe}$ and $N_{TeGe}$ at least partly compensate each other. The uncertainty of $N_{Ge}$ (and $N_{GeTe}$) is not symmetric: while fit qualities are significantly weaker upon decreasing coordination numbers by ~5%, the error is around 10% in the other direction.

The Ge-Te distance in Ge-Te glasses is 2.61 Å, which agrees with the value found in most telluride glasses (e.g. $Ge_{20}I_7Te_{73}$, $Ge_{11}Ga_{11}Te_{78}$ [8], Ge-As-Te [28], Ge-Te [29], Ge-Te-M (M: Cu, Ag, In) [6]). The peak is symmetric, without shoulder or secondary peak as in amorphous $Ge_2Sb_2Te_5$ [30, 31] or 0.75GeTe$_4$-0.25AgI [32].

The Te-Te distance is around 2.75 Å and does not depend significantly on the composition. The first peak of $g_{TeTe}(r)$ is well defined and followed by a deep, pronounced minimum.

### 4.3 Comparison with theoretical results

Short range order in Te-rich Ge-Te glasses has also been investigated by simulations based on density functional theory (DFT) [7, 33, 34]. Kalikka et al. used DFT with PBEsol exchange-correlation (XC) functional to optimize energetically model configurations obtained by fitting diffraction measurements with reverse Monte Carlo simulation [7].They found that Ge atoms are both in tetrahedral and defective octahedral configurations, in which Ge atoms have 3 nearest neighbors. The average total coordination number of Ge is 3.53 and the Ge-Te distance is 2.65 Å. Te atoms are mostly twofold coordinated ($N_{Te}$ = 2.13) and the Te-Te bond length is 2.84 Å.

The effect of XC functionals on the short range order in DFT models of glassy GeTe$_4$ was investigated by Bouzid et al [33]. It was concluded that structures obtained using BLYP functionals give a better agreement with X-ray diffraction data than those calculated with PBE functionals. It was also pointed out that van der Waals interactions should be taken into account to get reliable structural models with DFT.

The total coordination number of Ge is 3.97 and the Ge-Te distance is 2.59 Å in ref. [33]. Ge-Ge bonding was also found with a distance close to the value in Ge$_{23.6}$Te$_{76.4}$ and a coordination number around the experimentally detectable limit ($r_{GeGe}$ = 2.43 Å, $N_{GeGe}$ = 0.37]). The Te-Te distance is 2.84 Å while the total coordination number of Te atoms is 2.31.

It can thus be observed that by using BLYP functionals and taking into account van der Waals interactions the vast majority of Ge atoms is fourfold coordinated and the Ge-Ge and Ge-Te bond lengths are also close to the values of the present study and other experimental works combining diffraction and Ge EXAFS measurements [27, 28].

According to a recent ab initio molecular dynamics (AIMD) simulation using BLYP functionals [35] Ge atoms are mostly tetrahedrally coordinated in amorphous Ge$_2$Sb$_2$Te$_5$ but a significant fraction can be found in defective octahedral environment, which is characterized by longer Ge-Te bonds and smaller Te-Ge-Te angles. Longer Ge-Te bonds were also found in amorphous Ge$_2$Sb$_2$Te$_5$ using neutron diffraction, X-ray diffraction, EXAFS and reverse Monte Carlo simulation [30, 31]. The partly different nature of Ge-Te bonding in Ge-Te glasses and amorphous Ge$_2$Sb$_2$Te$_5$ is illustrated by Figure 5 where the first peaks of the Ge-Te partial pair correlation functions of amorphous Ge$_2$Sb$_2$Te$_5$ and Ge$_{18.7}$Te$_{81.3}$ are compared. The first peak of Ge$_{18.7}$Te$_{81.3}$ is well defined and symmetric while the shoulder clearly indicates the presence of longer Ge-Te bonds in Ge$_2$Sb$_2$Te$_5$.

These results show that under certain conditions AIMD simulations and experiment-based RMC modeling give a congruent description of the environment of Ge atoms. For AIMD the usage of BLYP functionals seems to be important while in case of RMC the moot point is the separation of the first peaks of Ge-Te and Te-Te partial pair correlation functions. Due to the strong overlap of these peaks X-ray and neutron diffraction data may not give sufficient information to avoid the admixture of Te-Te and Ge-Te peaks that may result in higher Ge-Te distance, decrease of Ge-Te coordination number and/or increase of the total coordination number of Te. This problem can be addressed by adding Ge EXAFS data that pinpoint the position of the Ge-Te peak and also help to decrease the uncertainty of coordination numbers.

The total coordination number of Te obtained by DFT studies is significantly higher than 2, regardless the type of XC functionals. The increased value is most likely due to the shallow first minimum of $g_{TeTe}(r)$ which suggests that the first and second coordination spheres of Te are not properly separated.

### 4.4. Corner sharing tetrahedra in Ge-Te glasses

For $x = 14.5$ and 18.7 the first peak of the Ge-Ge partial pair correlation function is located around 3.8 Å. It may be asked whether this well defined peak is due to the high Ge-Ge cut off distances (3.5 Å) or it originates from the particular information content of the experimental data. To decide this question we carried out a simulation of $Ge_{18.7}Te_{81.3}$ with lower cut off (3.0 Å) and another one in which neutron diffraction data were not fitted. The effect of these conditions on the first peak of $g_{GeGe}(r)$ is shown in Figure 6. The lower cut off has practically no influence on the peak position or area. On the other hand, the 'peak' obtained without neutron diffraction data is very flat and broad indicating that well defined peak shape is not a simulation artefact but is related directly with the neutron diffraction structure factor. Due to the higher neutron scattering length of Ge ($b_{Ge}$=8.185 fm, $b_{Te}$=5.80 fm [36]) neutron diffraction data are more sensitive to Ge-Ge correlations than X-ray diffraction, where the scattering power is determined essentially by the number of electrons ($Z_{Ge}$=32, $Z_{Te}$=52).

Even though the shape of Ge-Ge peak of interatomic separations is due to correlations between second neighbors, still it is well defined and highly symmetric. The sharpness of the peak is especially striking in comparison with the second peak of $g_{TeTe}(r)$. The latter describes correlations of Te atoms sharing common Ge *or* Te neighbors. A contribution from topologically distant (non-second neighbor) Te-Te pairs may also be significant. The small width of the Ge-Ge peak (~0.45 Å full width at half maximum) indicates that it is not composite in the above sense but rather corresponds to a single type correlation. The only reasonable choice is that the peak is due to Ge atoms sharing a common Te neighbor. As Ge is predominantly fourfold coordinated this most likely indicates the presence of *corner sharing GeTe₄ tetrahedra*. We note that with $r_{GeTe}$=2.60 Å and $r_{GeGe}$=3.80 Å one gets ~93° for the Ge-Te-Ge bond angle, which is a very reasonable value [7]. This assumption is supported by a constrained simulation run of $Ge_{18.7}Te_{81.3}$ in which each Ge atom was forced to have 4 Te neighbors while each Te was constrained to form bonds with 2 Ge or Te atoms. In addition, the Te-Ge-Te bond angle was constrained to be effectively in the 109.5° ± 15° range while the Ge-Te-Ge bond angle distribution was centered at 93°, with a similar spread. Detailed analysis of the resulting configuration revealed that the dominant contribution to the first peak of $g_{GeGe}(r)$ comes from Ge-Ge pairs centering corner sharing tetrahedra and there is a small fraction of edge sharing tetrahedra and pairs of topologically distant Ge atoms that do not have a common neighbor (Figure 7). The necessity of corner sharing was confirmed by a separate simulation run in which the above coordination and bond angle constraints were used but Ge-Te-Ge motifs were also forbidden in addition. It was found that neutron diffraction and Ge EXAFS fit residuals increased by ~100% and 65%, respectively, upon eliminating Te atoms with two Ge neighbors.

## Conclusions

Short range order in $Ge_xTe_{100-x}$ glasses ($x$ = 14.5, 18.7, 23.6) obtained by twin roller quenching technique was investigated by neutron diffraction, X-ray diffraction and Ge K-edge EXAFS measurements. Structural models were generated by fitting these data simultaneously by the reverse Monte Carlo simulation technique. By using unconstrained simulations it was established that Ge and Te atoms follow the 8-N rule. Ge-Ge bonds do not improve fit quality for $x$ = 14.5 and 18.7, indicating that these glasses cannot be described as random covalent networks. The presence of corner sharing $GeTe_4$ tetrahedra was inferred from the shape and position of the first peaks of the Ge-Ge partial pair correlation functions.

## Acknowledgements

The authors thank Mickaël Bigot for his help in sample preparation. Support from Agence Nationale de la Recherche (ANR) (Grant No. ANR-11-BS08-0012) is gratefully acknowledged. I. P. and P. J. were supported by NKFIH (National Research, Development and Innovation Office) Grant No. SNN 116198. The neutron diffraction experiment was carried out at the ORPHÉE reactor, Laboratoire Léon Brillouin, CEA-Saclay, France. We thank Diamond Light Source Ltd. for access to beamline I12-JEEP that contributed to the results presented here. EXAFS measurements were carried out at PETRA III at DESY, a member of the Helmholtz Association (HGF).

Table 1. Coordination numbers of atoms in Ge$_x$Te$_{100-x}$ glasses obtained by fitting neutron diffraction, X-ray diffraction and Ge K-edge EXAFS data by unconstrained reverse Monte Carlo simulations

|  | $N_{GeGe}$ | $N_{GeTe}$ | $N_{TeGe}$ | $N_{TeTe}$ | $N_{Ge}$ | $N_{Te}$ |
|---|---|---|---|---|---|---|
| Ge$_{14.5}$Te$_{85.5}$ | 0 | 4.16 (-0.2 + 0.6) | 0.73(-0.05 +0.1) | 1.31±0.15 | 4.16(-0.2 +0.6) | 2.04±0.1 |
| Ge$_{18.7}$Te$_{81.3}$ | 0 | 4.10 (-0.2 + 0.6) | 0.94(-0.06 +0.1) | 1.06±0.15 | 4.10(-0.2 +0.6) | 2.00±0.1 |
| Ge$_{23.6}$Te$_{76.4}$ | 0.44 ±0.3 | 3.66 (-0.2 + 0.6) | 1.13(-0.1 + 0.2) | 0.93±0.15 | 4.06(-0.2 +0.6) | 2.06±0.1 |

Table 2. Nearest neighbour interatomic distances in Ge$_x$Te$_{100-x}$ glasses obtained by fitting neutron diffraction, X-ray diffraction and Ge K-edge EXAFS data by applying unconstrained reverse Monte Carlo simulations

|  | $r_{GeGe}$ [Å] | $r_{GeTe}$ [Å] | $r_{TeTe}$ [Å] |
|---|---|---|---|
| Ge$_{14.5}$Te$_{85.5}$ | - | 2.61 ± 0.02 | 2.75 ± 0.02 |
| Ge$_{18.7}$Te$_{81.3}$ | - | 2.61 ± 0.02 | 2.76 ± 0.02 |
| Ge$_{23.6}$Te$_{76.4}$ | 2.45 ± 0.03 | 2.61 ± 0.02 | 2.75 ± 0.02 |

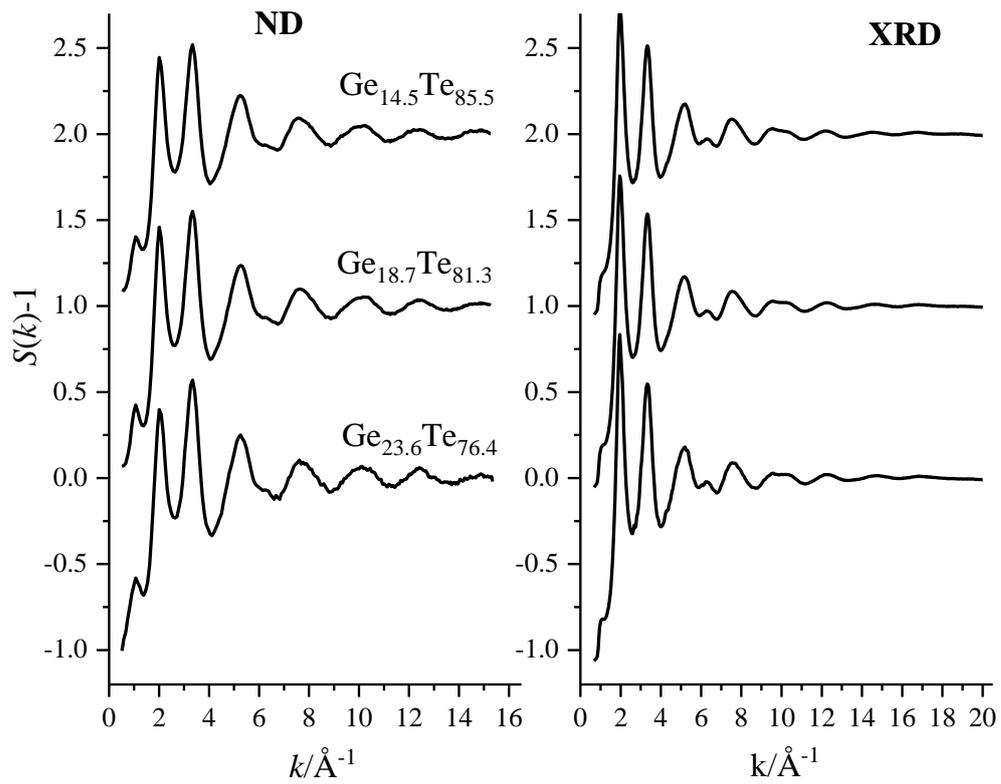

Figure 1. Neutron (ND) and X-ray (XRD) diffraction structure factors of the investigated Ge-Te glasses. Curves of $Ge_{18.7}Te_{81.3}$ and $Ge_{14.5}Te_{85.5}$ are shifted vertically with 1 and 2 units for clarity.

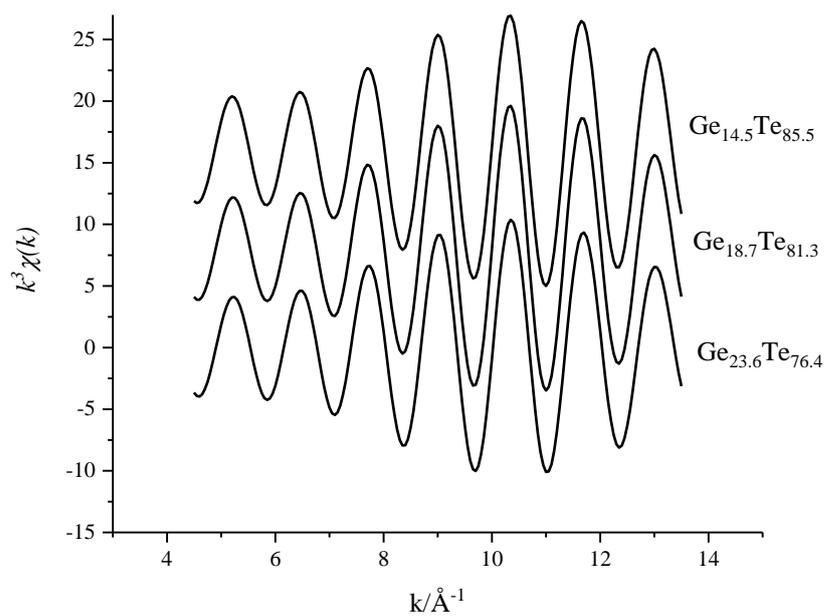

Figure 2. Filtered and $k^3$-weighted Ge K-edge EXAFS curves of the Ge-Te glasses investigated. Curves of $Ge_{18.7}Te_{81.3}$ and $Ge_{14.5}Te_{85.5}$ are shifted vertically with 7.5 and 15 units for clarity.

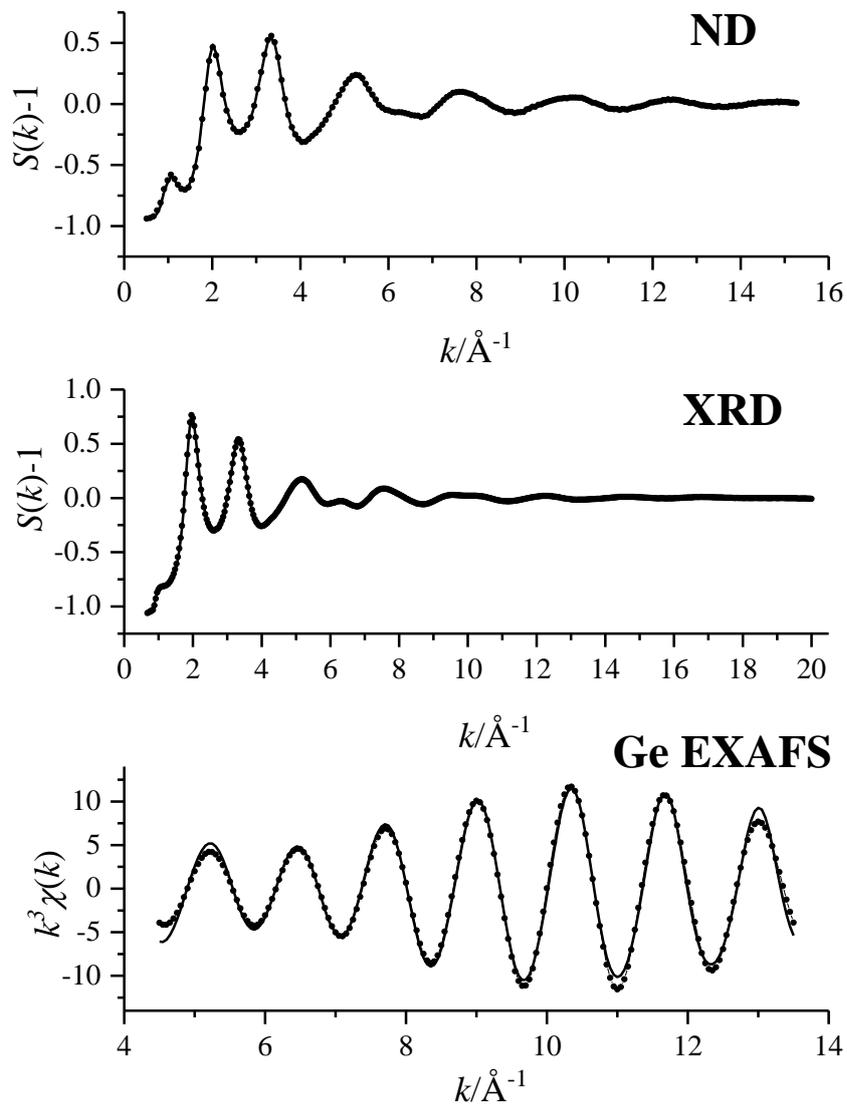

Figure 3. Comparison of experimental datasets (dotted line) and model curves (solid line) of Ge$_{18.7}$Te$_{81.3}$ obtained by fitting simultaneously the three measurements with reverse Monte Carlo simulation.

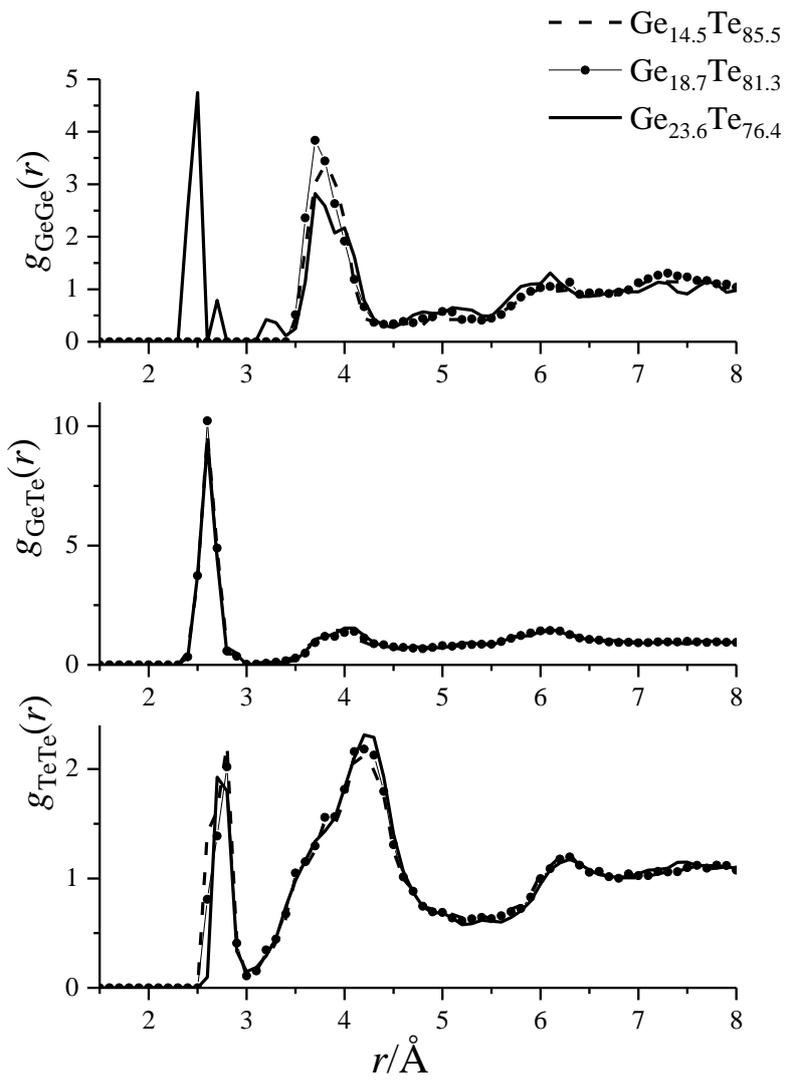

Figure 4. Partial pair correlation functions of the Ge-Te glasses obtained by reverse Monte Carlo simulation

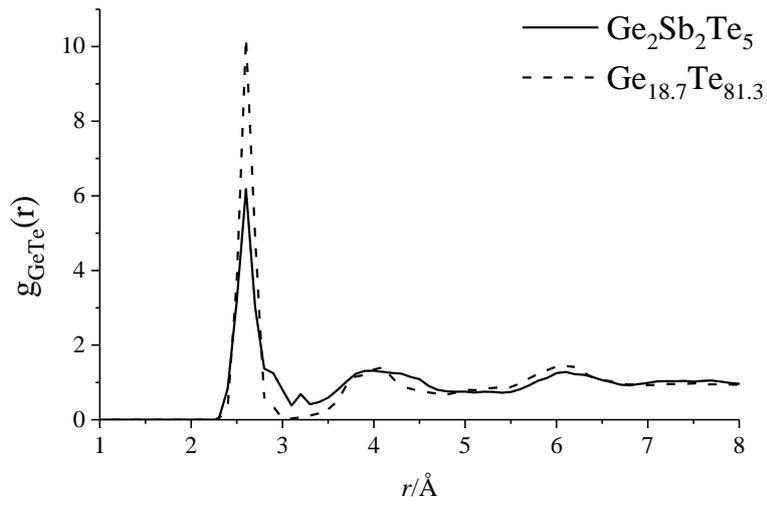

Figure 5. Ge-Te partial pair correlation functions of amorphous $Ge_2Sb_2Te_5$ [31] and glassy $Ge_{18.7}Te_{81.3}$ (this study)

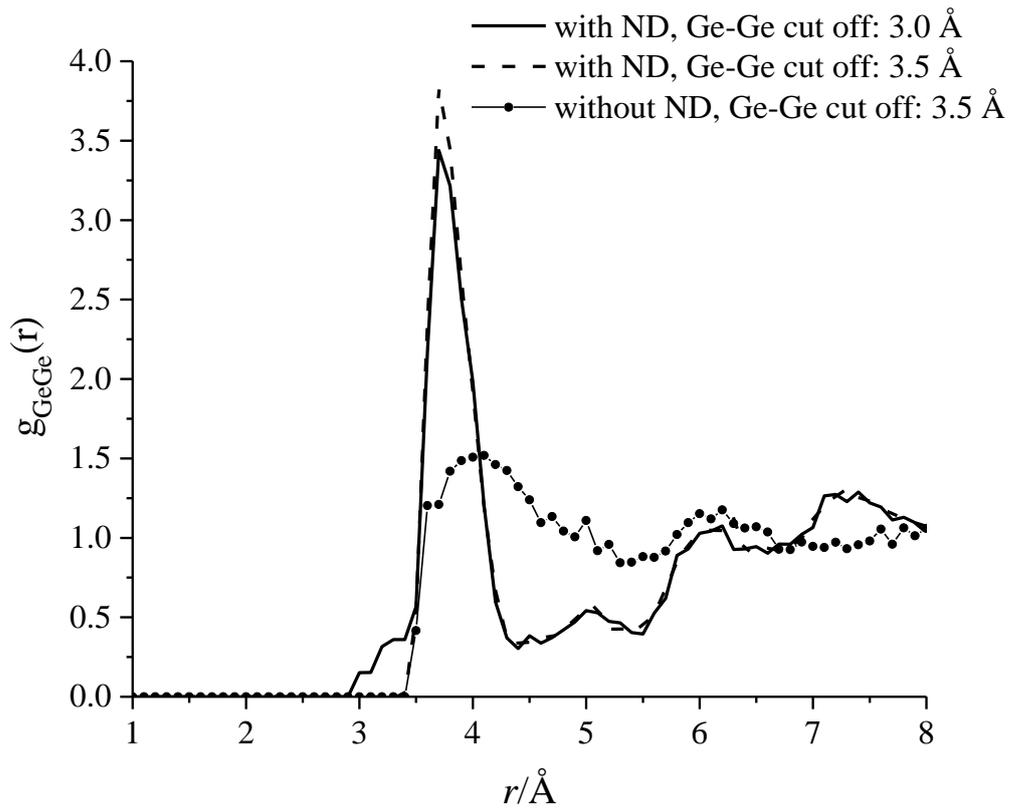

Figure 6. The first peak of $g_{GeGe}(r)$ of $Ge_{18.7}Te_{81.3}$ obtained with and without neutron diffraction data. The result of a simulation with neutron diffraction data but lower minimum Ge-Ge distance is also shown

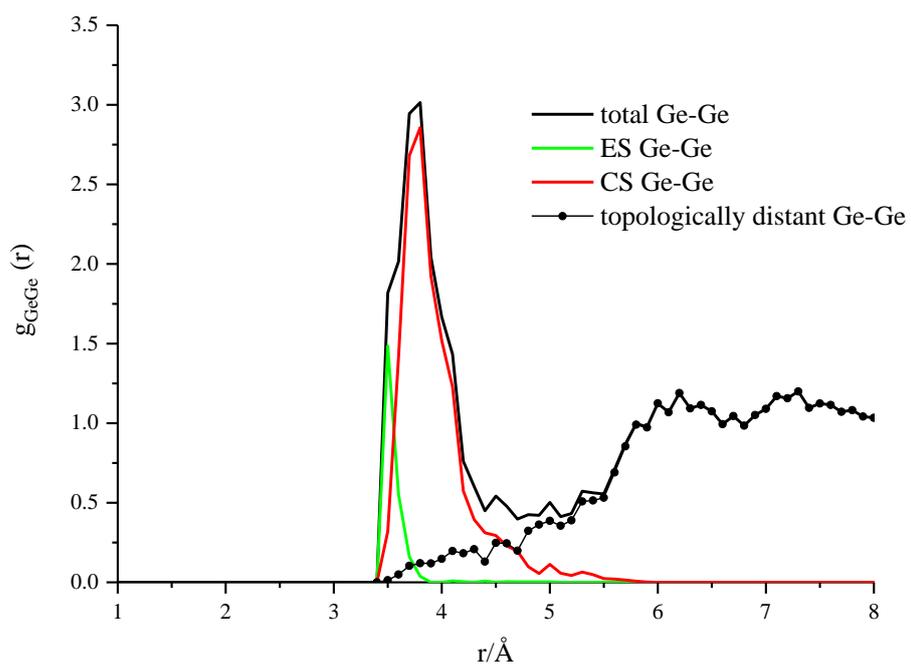

Figure 7. Decomposition of the first peak of $g_{GeGe}(r)$ of $Ge_{18.7}Te_{81.3}$ to contributions from corner sharing (CS) tetrahedra, edge sharing (ES) tetrahedra and topologically distant Ge-Ge pairs